# Fault tolerant authenticated quantum dialogue using logical Bell states

Tian-Yu Ye*

College of Information & Electronic Engineering, Zhejiang Gongshang University, Hangzhou 310018, P.R.China

**Abstract**
Two fault tolerant authenticated quantum dialogue (AQD) protocols are proposed in this paper by employing logical Bell states as the quantum resource, which combat the collective-dephasing noise and the collective-rotation noise, respectively. The two proposed protocols each can accomplish the mutual identity authentication and the dialogue between two participants simultaneously and securely over one kind of collective noise channels. In each of two proposed protocols, the information transmitted through the classical channel is assumed to be eavesdroppable and modifiable. The key for choosing the measurement bases of sample logical qubits is pre-shared privately between two participants. The Bell state measurements rather than the four-qubit joint measurements are adopted for decoding. The two participants share the initial states of message logical Bell states with resort to the direct transmission of auxiliary logical Bell states so that the information leakage problem is avoided. The impersonation attack, the man-in-the-middle attack, the modification attack and the Trojan horse attacks from Eve all are detectable.
**Keywords:** Authenticated quantum dialogue (AQD), collective noise, logical Bell state, information leakage, impersonation attack, man-in-the-middle attack, modification attack, Trojan horse attack.

## 1 Introduction

In 2002, Long and Liu [1] proposed the first quantum secure direct communication (QSDC) protocol called the state-encoded two-step protocol, which realizes the direct transmission of secret messages between two participants through the transmission of quantum signal for the first time. This protocol also invented the important idea of quantum data block transmission. Since then, numerous good QSDC protocols have been constructed [2-11].

Later, in 2004, Nguyen [12] and Zhang et al. [13-14] independently proposed the concept named quantum dialogue (QD), which is able to realize the bidirectional communication of secret messages between two participants through the transmission of quantum signal. Afterward, a lot of QD protocols were suggested with different quantum technologies [15-24]. Unfortunately, they always run the security risk named information leakage, which was firstly discovered by Gao et al. [25-26] in 2008. Essentially, the information leakage problem of QD is derived from the phenomenon of classical correlation which was pointed out by Tan and Cai [27] in 2008. Subsequently, how to design the QD protocols without information leakage quickly aroused the interests of researchers. As a result, many good information leakage resistant QD protocols have been designed [28-38].

However, all the above QD protocols [12-24,28-38] only work on the basis that the information transmitted through the classical channel is not eavesdroppable and modifiable. Once this assumption is broken, they immediately face with serious security risks aroused by the attacks from an outside attacker, such as the impersonation attack, the man-in-the-middle attack and the modification attack. Fortunately, a special kind of QD called authenticated QD (AQD) has been suggested by Naseri [39] and Shen et al. [40] to make up for this drawback. However, in 2013, Lin et al. [41] pointed out that Shen et al.'s protocol [40] is threatened by the man-in-the-middle attack. There is no exception for Naseri's protocol [39] under this type of attack. Subsequently, in order to improve the security of AQD, Lin et al. [42] proposed a secure AQD with Bell states, where two participants are able to accomplish the mutual identity authentication and the dialogue simultaneously and securely. In Lin et al.'s protocol [42], two kinds of keys are pre-shared privately between two participants, where the first one is used for preparing the initial Bell states. In this sense, two participants pre-share the initial Bell states. However, in the realm of information leakage resistant QD, the initial states of quantum states are always not assumed to be pre-shared between two participants. That is to say, two participants have to resort to a certain kind of quantum technology to share the initial states of quantum states. Another defect of Lin et al.'s protocol [42] is that it only works over an ideal channel. It is necessary to point out that the previous information leakage resistant QD protocols [28-38] each is also only workable over an ideal channel. Actually, there inevitably exists the unwanted coupling of photons with the practical environment. Consequently, the influence of channel noise cannot be ignored. As pointed out in Refs.[43-44], the channel noise can be regarded to be collective on the basis that photons travel inside a time window which is shorter than the variation of noise. There are usually two kinds of collective noise. The first one is the collective-dephasing noise, which always keeps the horizontal polarization state of photon $|0\rangle$ unchanged and converts the vertical polarization state of photon $|1\rangle$ into $e^{i\varphi}|1\rangle$ [45]. The other one is the collective-rotation noise, which always converts $|0\rangle$ and $|1\rangle$ into $\cos\theta|0\rangle+\sin\theta|1\rangle$ and $-\sin\theta|0\rangle+\cos\theta|1\rangle$, respectively [47]. Here, $\varphi$ and $\theta$ are the corresponding noise parameters fluctuating with time. Decoherence-free (DF) states [43-61], which are invariant against the collective noise, have been a kind of effective means to combat the collective noise so that they have been frequently adopted to design antinoise quantum cryptography protocols. For example, the robust QD protocols in Ref.[53] have used the product states of two-qubit DF states as the quantum resource; the robust QD protocols in Refs.[54-55,57] have used the two-qubit DF states as the quantum resource; the robust QD protocols in Ref.[61] have used the three-qubit DF states as the quantum resource; and the robust QD protocols in Refs.[56,58-60] have used the four-qubit DF states as the quantum resource.

---

*Corresponding author:
 E-mail：happyyty@aliyun.com(T.Y.Ye)

Based on the above analysis, in this paper, two fault tolerant AQD protocols are proposed by employing logical Bell states (a kind of four-qubit DF states) as the quantum resource, where the information transmitted through the classical channel is assumed to be eavesdroppable and modifiable. The two proposed protocols are constructed against the collective-dephasing noise and the collective-rotation noise, respectively. Each of them can accomplish the mutual identity authentication and the dialogue between two participants simultaneously and securely over one kind of collective noise channels. Similar to the protocol of Ref.[42], the key for choosing the measurement bases of sample logical qubits is pre-shared privately between two participants. The Bell state measurements rather than the four-qubit joint measurements are adopted for decoding. Similar to the QD protocols of Ref.[56], the two participants share the initial states of message logical Bell states with resort to the direct transmission of auxiliary logical Bell states. In this way, the information leakage problem is avoided. The impersonation attack, the man-in-the-middle attack and the modification attack and the Trojan horse attacks from Eve all are detectable.

## 2 Fault tolerant AQD protocol against the collective-dephasing noise

The four logical Bell states under the collective-dephasing noise can be expressed as [49]

$$|\Phi_{dp}^+\rangle_{1234} = \frac{1}{\sqrt{2}}(|0_{dp}\rangle|0_{dp}\rangle + |1_{dp}\rangle|1_{dp}\rangle)_{1234} = \frac{1}{\sqrt{2}}(|01\rangle|01\rangle + |10\rangle|10\rangle)_{1234} = \frac{1}{\sqrt{2}}(|+_{dp}\rangle|+_{dp}\rangle + |-_{dp}\rangle|-_{dp}\rangle)_{1234}$$
$$= \frac{1}{\sqrt{2}}(|00\rangle|11\rangle + |11\rangle|00\rangle)_{1324} = \frac{1}{\sqrt{2}}(|\phi^+\rangle|\phi^+\rangle - |\phi^-\rangle|\phi^-\rangle)_{1324},$$

$$|\Phi_{dp}^-\rangle_{1234} = \frac{1}{\sqrt{2}}(|0_{dp}\rangle|0_{dp}\rangle - |1_{dp}\rangle|1_{dp}\rangle)_{1234} = \frac{1}{\sqrt{2}}(|01\rangle|01\rangle - |10\rangle|10\rangle)_{1234} = \frac{1}{\sqrt{2}}(|+_{dp}\rangle|-_{dp}\rangle + |-_{dp}\rangle|+_{dp}\rangle)_{1234}$$
$$= \frac{1}{\sqrt{2}}(|00\rangle|11\rangle - |11\rangle|00\rangle)_{1324} = \frac{1}{\sqrt{2}}(|\phi^-\rangle|\phi^+\rangle - |\phi^+\rangle|\phi^-\rangle)_{1324},$$

$$|\Psi_{dp}^+\rangle_{1234} = \frac{1}{\sqrt{2}}(|0_{dp}\rangle|1_{dp}\rangle + |1_{dp}\rangle|0_{dp}\rangle)_{1234} = \frac{1}{\sqrt{2}}(|01\rangle|10\rangle + |10\rangle|01\rangle)_{1234} = \frac{1}{\sqrt{2}}(|+_{dp}\rangle|+_{dp}\rangle - |-_{dp}\rangle|-_{dp}\rangle)_{1234}$$
$$= \frac{1}{\sqrt{2}}(|01\rangle|10\rangle + |10\rangle|01\rangle)_{1324} = \frac{1}{\sqrt{2}}(|\psi^+\rangle|\psi^+\rangle - |\psi^-\rangle|\psi^-\rangle)_{1324},$$

$$|\Psi_{dp}^-\rangle_{1234} = \frac{1}{\sqrt{2}}(|0_{dp}\rangle|1_{dp}\rangle - |1_{dp}\rangle|0_{dp}\rangle)_{1234} = \frac{1}{\sqrt{2}}(|01\rangle|10\rangle - |10\rangle|01\rangle)_{1234} = \frac{1}{\sqrt{2}}(|+_{dp}\rangle|-_{dp}\rangle - |-_{dp}\rangle|+_{dp}\rangle)_{1234}$$
$$= \frac{1}{\sqrt{2}}(|01\rangle|10\rangle - |10\rangle|01\rangle)_{1324} = \frac{1}{\sqrt{2}}(|\psi^-\rangle|\psi^+\rangle - |\psi^+\rangle|\psi^-\rangle)_{1324}, \tag{1}$$

where $|\phi^\pm\rangle = \frac{1}{\sqrt{2}}(|00\rangle \pm |11\rangle)$ and $|\psi^\pm\rangle = \frac{1}{\sqrt{2}}(|01\rangle \pm |10\rangle)$ are four original Bell states. Here, $|0_{dp}\rangle = |01\rangle$ and $|1_{dp}\rangle = |10\rangle$ are two logical qubits immune to the collective-dephasing noise [45]. Moreover, $|\pm_{dp}\rangle = \frac{1}{\sqrt{2}}(|0_{dp}\rangle \pm |1_{dp}\rangle) = \frac{1}{\sqrt{2}}(|01\rangle \pm |10\rangle)$ are their superpositions, which are also invariant against this kind of noise [47]. Accordingly, two logical measuring bases under this kind of noise, i.e., $Z_{dp} = \{|0_{dp}\rangle, |1_{dp}\rangle\}$ and $X_{dp} = \{|+_{dp}\rangle, |-_{dp}\rangle\}$, form. It is easy to know from Eq.(1) that as long as two Bell state measurements are imposed on the 1st and the 3rd qubits and on the 2nd and the 4th qubits, respectively, the above four logical Bell states can be distinguished from each other [49]. On the other hand, as summarized in Table 1, these four logical Bell states can be mutually converted through the four logical unitary operations under this kind of noise [49-50], which are defined as

$$\Omega_I = I_1 \otimes I_2, \Omega_z = U_{z1} \otimes I_2, \Omega_x = U_{x1} \otimes U_{x2}, \Omega_y = U_{y1} \otimes U_{x2}. \tag{2}$$

Here, $I = |0\rangle\langle 0| + |1\rangle\langle 1|$, $U_x = |1\rangle\langle 0| + |0\rangle\langle 1|$, $U_y = |0\rangle\langle 1| - |1\rangle\langle 0|$ and $U_z = |0\rangle\langle 0| - |1\rangle\langle 1|$.

Table 1  Dense coding of the four logical Bell states under the collective-dephasing noise

|  | $|\Phi_{dp}^+\rangle$ | $|\Phi_{dp}^-\rangle$ | $|\Psi_{dp}^+\rangle$ | $|\Psi_{dp}^-\rangle$ |
|---|---|---|---|---|
| $\Omega_I$ | $|\Phi_{dp}^+\rangle$ | $|\Phi_{dp}^-\rangle$ | $|\Psi_{dp}^+\rangle$ | $|\Psi_{dp}^-\rangle$ |
| $\Omega_z$ | $|\Phi_{dp}^-\rangle$ | $|\Phi_{dp}^+\rangle$ | $|\Psi_{dp}^-\rangle$ | $|\Psi_{dp}^+\rangle$ |
| $\Omega_x$ | $|\Psi_{dp}^+\rangle$ | $|\Psi_{dp}^-\rangle$ | $|\Phi_{dp}^+\rangle$ | $|\Phi_{dp}^-\rangle$ |
| $\Omega_y$ | $|\Psi_{dp}^-\rangle$ | $|\Psi_{dp}^+\rangle$ | $|\Phi_{dp}^-\rangle$ | $|\Phi_{dp}^+\rangle$ |

Suppose that Alice and Bob agree on in advance that the four logical unitary operations described in Eq.(2) represent the classical bits in such a way that

$$\Omega_I \to \Omega_{00}, \Omega_z \to \Omega_{01}, \Omega_x \to \Omega_{10}, \Omega_y \to \Omega_{11}. \qquad (3)$$

Here, each digital subscript denotes a classical two-bit. On the other hand, suppose that Alice has $3N/2$ classical bits $\{(i_1,j_1)(i_2,j_2)\cdots(i_d,j_d)\cdots(i_{3N/4},j_{3N/4})\}$ which are denoted as $m_A \| h(m_A)$. Here, $m_A$ and $h(m_A)$ are her secret and its corresponding hash value, respectively, and "$\|$" is the sign of concatenation. Similarly, suppose that Bob has $3N/2$ classical bits $\{(t_1,l_1)(t_2,l_2)\cdots(t_d,l_d)\cdots(t_{3N/4},l_{3N/4})\}$ which are denoted as $m_B \| h(m_B)$. Here, $m_B$ is his secret and $h(m_B)$ is its corresponding hash value. Apparently, $i_d, j_d, t_d, l_d \in \{0,1\}$, where $d \in \{1,2,\cdots,3N/4\}$. Similar to the protocol of Ref.[42], one key $K$ is further assumed to be pre-shared privately between Alice and Bob, which is used for choosing the measurement bases of sample logical qubits here. The length of $K$ is $N/4$ so that $K \in \{0,1\}^{N/4}$. Concretely speaking, if the $p^{th}$ $(p=1,2,\cdots,N/4)$ bit of $K$ is 0, the base $Z_{dp}$ is adopted to measure the $p^{th}$ sample logical qubit; otherwise, the base $X_{dp}$ is employed to measure it.

The AQD protocol against the collective-dephasing noise is made up of the following steps. Here, the information transmitted through the classical channel is assumed to be eavesdroppable and modifiable.

**Step 1:** Alice prepares $2N$ message logical Bell states $\{(A_1,B_1),(A_2,B_2),\cdots,(A_{2n-1},B_{2n-1}),(A_{2n},B_{2n}),\cdots,(A_{2N-1},B_{2N-1}),(A_{2N},B_{2N})\}$ $(n=1,2,\cdots,N)$, where each two adjacent states $(A_{2n-1},B_{2n-1})$ and $(A_{2n},B_{2n})$ are made in the same state (randomly in one of the four states $\{|\Phi_{dp}^+\rangle,|\Phi_{dp}^-\rangle,|\Psi_{dp}^+\rangle,|\Psi_{dp}^-\rangle\}$), similar to the first protocol of Ref.[56]. Alice picks out the logical qubit $A$ with odd subscripts and even subscripts to make up sequences $S_{OA}$ and $S_{EA}$, respectively. That is to say, $S_{OA}=\{A_1,A_3,\cdots,A_{2n-1},\cdots,A_{2N-1}\}$ and $S_{EA}=\{A_2,A_4,\cdots,A_{2n},\cdots,A_{2N}\}$. Alice does the same thing on the logical qubit $B$. Accordingly, it follows that $S_{OB}=\{B_1,B_3,\cdots,B_{2n-1},\cdots,B_{2N-1}\}$ and $S_{EB}=\{B_2,B_4,\cdots,B_{2n},\cdots,B_{2N}\}$. Then, for security check and mutual identity authentication, Alice prepares $N/4$ sample logical Bell states and divides them into two subsequences composed by the sample logical qubits $A$ and $B$, respectively. Afterward, Alice randomly inserts the subsequence with the sample logical qubit $A$ into $S_{EA}$ to form $S'_{EA}$. Likewise, Alice does the similar thing on the subsequence with the sample logical qubit $B$ and $S_{EB}$ to form $S'_{EB}$. Finally, Alice keeps $S_{OA}$, $S'_{EA}$ and $S_{OB}$ intact, and sends $S'_{EB}$ to Bob via a quantum channel. Apparently, $S'_{EB}$ is sent out in a block transmission manner [1].

**Step 2:** After Bob informs Alice of his receipt of $S'_{EB}$, Alice checks its transmission security and authenticates the identity of Bob. Alice tells Bob the positions of the sample logical qubits in $S'_{EB}$ via a classical channel. According to $K$, Bob measures the sample logical qubits in $S'_{EB}$ and obtains their corresponding measurement result $R_{S_{EB}}$. Then, Bob sends $R_{S_{EB}}$ to Alice via the classical channel. According to $K$, Alice measures the sample logical qubits in $S'_{EA}$ and obtains their corresponding measurement result $R_{S_{EA}}$. As Alice prepares the sample logical Bell states by herself, according to $R_{S_{EB}}$ and $R_{S_{EA}}$, she can judge out the existence of an eavesdropper during the transmission of $S'_{EB}$ and the identity of Bob simultaneously. If something wrong is discovered by Alice, she terminates the communication; otherwise, she sends $S'_{EA}$ to Bob via the quantum channel also in a block transmission manner [1].

**Step 3:** After Bob informs Alice of his receipt of $S'_{EA}$, Alice tells Bob the positions of the sample logical qubits in $S'_{EA}$ via the classical channel. Then, Bob discards the sample logical qubits in both $S'_{EA}$ and $S'_{EB}$. Accordingly, $S'_{EA}$ and $S'_{EB}$ are turned back into $S_{EA}$ and $S_{EB}$, respectively. Afterward, Bob picks the logical qubits $A_{2n}$ and $B_{2n}$ out from $S_{EA}$ and $S_{EB}$, respectively, and restores them as the $2n^{th}$ message logical Bell state. Then, in order to know the state of $(A_{2n},B_{2n})$, Bob performs two Bell state measurements on the 1st and the 3rd qubits and on the 2nd and the 4th qubits, respectively. As each two adjacent message logical Bell states $(A_{2n-1},B_{2n-1})$ and $(A_{2n},B_{2n})$ are made in the same state by Alice in Step 1, Bob can directly know the state of $(A_{2n-1},B_{2n-1})$ from that of $(A_{2n},B_{2n})$.

**Step 4:** Alice sends $S_{OB}$ to Bob via the quantum channel also in a block transmission manner [1]. Then Bob chooses $N/4$ logical qubits from $S_{OB}$ as the sample logical qubits and tells Alice their positions via the classical channel. According to $K$, Alice measures the corresponding sample logical qubits in $S_{OA}$ and obtains their corresponding measurement result $R_{S_{OA}}$. Then, Alice sends $R_{S_{OA}}$ to Bob via the classical channel. According to $K$, Bob measures the sample logical qubits in $S_{OB}$ and obtains their corresponding measurement result $R_{S_{OB}}$. As Bob knows the state of $(A_{2n-1}, B_{2n-1})$ in Step 3, according to $R_{S_{OA}}$ and $R_{S_{OB}}$, he can judge out the existence of an eavesdropper during the transmission of $S_{OB}$ and the identity of Alice simultaneously. If something wrong is discovered by Bob, he terminates the communication; otherwise, the communication is continued.

**Step 5:** Alice and Bob discard the sample logical qubits in $S_{OA}$ and $S_{OB}$ in order, respectively. For convenience, the remaining $S_{OA}$ and $S_{OB}$ are represented by $S'_{OA} = \{A'_1, A'_2, \cdots, A'_{3N/4}\}$ and $S'_{OB} = \{B'_1, B'_2, \cdots, B'_{3N/4}\}$, respectively. Then, Alice performs the logical unitary operation $\Omega_{i_d j_d}$ on $A'_d$ $(d=1,2,\cdots,3N/4)$ for encoding. As a result, $S'_{OA}$ is changed into $S''_{OA} = \{\Omega_{i_1 j_1} A'_1, \Omega_{i_2 j_2} A'_2, \cdots, \Omega_{i_{3N/4} j_{3N/4}} A'_{3N/4}\}$. Finally, Alice sends $S''_{OA}$ to Bob via the quantum channel also in a block transmission manner [1].

**Step 6:** After receiving $S''_{OA}$, Bob performs the logical unitary operation $\Omega_{t_d l_d}$ on $\Omega_{i_d j_d} A'_d$ for encoding. As a result, $S''_{OA}$ is changed into $S'''_{OA} = \{\Omega_{t_1 l_1}\Omega_{i_1 j_1} A'_1, \Omega_{t_2 l_2}\Omega_{i_2 j_2} A'_2, \cdots, \Omega_{t_{3N/4} l_{3N/4}}\Omega_{i_{3N/4} j_{3N/4}} A'_{3N/4}\}$. Afterward, Bob picks $\Omega_{t_d l_d}\Omega_{i_d j_d} A'_d$ and $B'_d$ out from $S'''_{OA}$ and $S'_{OB}$, respectively, and makes them form the $d^{\text{th}}$ $(d \in \{1,2,\cdots,3N/4\})$ group. Then, in order to know the state of $(\Omega_{t_d l_d}\Omega_{i_d j_d} A'_d, B'_d)$, Bob performs two Bell state measurements on the 1st and the 3rd qubits and on the 2nd and the 4th qubits, respectively. According to his own logical unitary operation $\Omega_{t_d l_d}$ and his own measurement result of $(\Omega_{t_d l_d}\Omega_{i_d j_d} A'_d, B'_d)$, Bob is able to read out $(i_d, j_d)$, as he knows the state of $(A'_d, B'_d)$. Here, the decoded classical bits are denoted as $m'_A \| h(m_A)'$, where $m'_A$ and $h(m_A)'$ are the decoded versions of $m_A$ and $h(m_A)$, respectively. Then, Bob computes the hash value of $m'_A$ called $h(m'_A)$ and compares $h(m'_A)$ with $h(m_A)'$, similar to the protocol of Ref.[42]. If they are identical, $m'_A$ is regarded as the genuine $m_A$ by Bob so that the communication is continued; otherwise, the communication is halted.

**Step 7:** Bob sends his own measurement result of $\{(\Omega_{t_1 l_1}\Omega_{i_1 j_1} A'_1, B'_1), (\Omega_{t_2 l_2}\Omega_{i_2 j_2} A'_2, B'_2), \cdots, (\Omega_{t_{3N/4} l_{3N/4}}\Omega_{i_{3N/4} j_{3N/4}} A'_{3N/4}, B'_{3N/4})\}$ to Alice via the classical channel. Likewise, Alice is able to know $(t_d, l_d)$ $(d \in \{1,2,\cdots,3N/4\})$ according to her own logical unitary operation $\Omega_{i_d j_d}$, as she prepares $(A'_d, B'_d)$ by herself. Here, the decoded classical bits are denoted as $m'_B \| h(m_B)'$, where $m'_B$ and $h(m_B)'$ are the decoded versions of $m_B$ and $h(m_B)$, respectively. Then, Alice computes the hash value of $m'_B$ called $h(m'_B)$ and compares $h(m'_B)$ with $h(m_B)'$, similar to the protocol of Ref.[42]. If they are identical, $m'_B$ is regarded as the genuine $m_B$ by Alice so that the communication is completely successful; otherwise, the communication fails and starts from the beginning.

After generalizing the protocol of Ref.[42] into the case of collective-dephasing noise by using the logical Bell states in Eq.(1) to replace the original Bell states used in Ref.[42], it is easy to find out that the above protocol is highly related to both the first protocol of Ref.[56] and the collective-dephasing noise version of the protocol in Ref.[42]. In fact, the above protocol can be regarded as the combination of these two protocols.

## 3 Fault tolerant AQD protocol against the collective-rotation noise

The four logical Bell states under the collective-rotation noise can be expressed as [49]

$$|\Phi_r^+\rangle_{1234} = \frac{1}{\sqrt{2}}(|0_r\rangle|0_r\rangle + |1_r\rangle|1_r\rangle)_{1234} = \frac{1}{\sqrt{2}}(|\phi^+\rangle|\phi^+\rangle + |\psi^-\rangle|\psi^-\rangle)_{1234}$$

$$= \frac{1}{\sqrt{2}}(|+_r\rangle|+_r\rangle + |-_r\rangle|-_r\rangle)_{1234} = \frac{1}{\sqrt{2}}(|\phi^+\rangle|\phi^+\rangle + |\psi^-\rangle|\psi^-\rangle)_{1324},$$

$$|\Phi_r^-\rangle_{1234} = \frac{1}{\sqrt{2}}(|0_r\rangle|0_r\rangle - |1_r\rangle|1_r\rangle)_{1234} = \frac{1}{\sqrt{2}}(|\phi^+\rangle|\phi^+\rangle - |\psi^-\rangle|\psi^-\rangle)_{1234}$$

$$= \frac{1}{\sqrt{2}}(|+_r\rangle|-_r\rangle + |-_r\rangle|+_r\rangle)_{1234} = \frac{1}{\sqrt{2}}(|\phi^-\rangle|\phi^-\rangle + |\psi^+\rangle|\psi^+\rangle)_{1324},$$

$$|\Psi_r^+\rangle_{1234} = \frac{1}{\sqrt{2}}(|0_r\rangle|1_r\rangle + |1_r\rangle|0_r\rangle)_{1234} = \frac{1}{\sqrt{2}}(|\phi^+\rangle|\psi^-\rangle + |\psi^-\rangle|\phi^+\rangle)_{1234}$$

$$= \frac{1}{\sqrt{2}}(|+_r\rangle|+_r\rangle - |-_r\rangle|-_r\rangle)_{1234} = \frac{1}{\sqrt{2}}(|\phi^-\rangle|\psi^+\rangle - |\psi^+\rangle|\phi^-\rangle)_{1324},$$

$$|\Psi_r^-\rangle_{1234} = \frac{1}{\sqrt{2}}(|0_r\rangle|1_r\rangle - |1_r\rangle|0_r\rangle)_{1234} = \frac{1}{\sqrt{2}}(|\phi^+\rangle|\psi^-\rangle - |\psi^-\rangle|\phi^+\rangle)_{1234}$$

$$= \frac{1}{\sqrt{2}}(|+_r\rangle|-_r\rangle - |-_r\rangle|+_r\rangle)_{1234} = \frac{1}{\sqrt{2}}(|\phi^+\rangle|\psi^-\rangle - |\psi^-\rangle|\phi^+\rangle)_{1324}. \quad (4)$$

Here, $|0_r\rangle = |\phi^+\rangle$ and $|1_r\rangle = |\psi^-\rangle$ are two logical qubits immune to the collective-rotation noise [47]. Moreover, $|\pm_r\rangle = \frac{1}{\sqrt{2}}(|0_r\rangle \pm |1_r\rangle) = \frac{1}{\sqrt{2}}(|\phi^+\rangle \pm |\psi^-\rangle)$ are their superpositions, which are also invariant against this kind of noise [49]. Accordingly, two logical measuring bases under this kind of noise, i.e., $Z_r = \{|0_r\rangle, |1_r\rangle\}$ and $X_r = \{|+_r\rangle, |-_r\rangle\}$, form. It is easy to know from Eq.(4) that as long as two Bell state measurements are imposed on the 1st and the 3rd qubits and on the 2nd and the 4th qubits, respectively, the above four logical Bell states can be distinguished from each other [49]. On the other hand, as summarized in Table 2, these four logical Bell states can be mutually converted through the four logical unitary operations under this kind of noise [49-50], which are defined as

$$\Theta_I = I_1 \otimes I_2, \Theta_z = U_{z1} \otimes U_{z2}, \Theta_x = U_{z1} \otimes U_{x2}, \Theta_y = I_1 \otimes U_{y2}. \quad (5)$$

Table 2 Dense coding of the four logical Bell states under the collective-rotation noise

|  | $|\Phi_r^+\rangle$ | $|\Phi_r^-\rangle$ | $|\Psi_r^+\rangle$ | $|\Psi_r^-\rangle$ |
|---|---|---|---|---|
| $\Theta_I$ | $|\Phi_r^+\rangle$ | $|\Phi_r^-\rangle$ | $|\Psi_r^+\rangle$ | $|\Psi_r^-\rangle$ |
| $\Theta_z$ | $|\Phi_r^-\rangle$ | $|\Phi_r^+\rangle$ | $|\Psi_r^-\rangle$ | $|\Psi_r^+\rangle$ |
| $\Theta_x$ | $|\Psi_r^+\rangle$ | $|\Psi_r^-\rangle$ | $|\Phi_r^+\rangle$ | $|\Phi_r^-\rangle$ |
| $\Theta_y$ | $|\Psi_r^-\rangle$ | $|\Psi_r^+\rangle$ | $|\Phi_r^-\rangle$ | $|\Phi_r^+\rangle$ |

Suppose that Alice and Bob agree on in advance that the four logical unitary operations described in Eq.(5) represent the classical bits in such a way that

$$\Theta_I \to \Theta_{00}, \Theta_z \to \Theta_{01}, \Theta_x \to \Theta_{10}, \Theta_y \to \Theta_{11}. \quad (6)$$

Here, each digital subscript also denotes a classical two-bit. Same to the protocol in Section 2, suppose that $m_A \| h(m_A)$ denotes Alice's $3N/2$ classical bits $\{(i_1, j_1)(i_2, j_2)\cdots(i_d, j_d)\cdots(i_{3N/4}, j_{3N/4})\}$, and $m_B \| h(m_B)$ denotes Bob's $3N/2$ classical bits $\{(t_1, l_1)(t_2, l_2)\cdots(t_d, l_d)\cdots(t_{3N/4}, l_{3N/4})\}$. Here, $i_d, j_d, t_d, l_d \in \{0,1\}$, where $d \in \{1, 2, \cdots, 3N/4\}$. Similar to the protocol of Ref.[42], one key $K$ is further assumed to be pre-shared privately between Alice and Bob, which is used for choosing the measurement bases of sample logical qubits here. The length of $K$ is $N/4$ so that $K \in \{0,1\}^{N/4}$. Concretely speaking, if the $p^{\text{th}}$ $(p = 1, 2, \cdots, N/4)$ bit of $K$ is 0, the base $Z_r$ is adopted to measure the $p^{\text{th}}$ sample logical qubit; otherwise, the base $X_r$ is employed to measure it.

The AQD protocol against the collective-dephasing noise in Section 2 can be immediately changed into the one against the collective-rotation noise, as long as the following modifications are made.

(1) In Step 1, each two adjacent logical Bell states $(A_{2n-1}, B_{2n-1})$ and $(A_{2n}, B_{2n})$ $(n = 1, 2, \cdots, N)$ are prepared by Alice in the same state randomly in one of the four states $\{|\Phi_r^+\rangle, |\Phi_r^-\rangle, |\Psi_r^+\rangle, |\Psi_r^-\rangle\}$, similar to the second protocol of Ref.[56];

(2) In Step 5, the logical unitary operation $\Theta_{i_d j_d}$ is performed by Alice on $A_d^{'}$ $(d = 1, 2, \cdots, 3N/4)$ for encoding;

(3) In Step 6, the logical unitary operation $\Theta_{t_d l_d}$ is performed by Bob on $\Theta_{i_d j_d} A_d^{'}$ for encoding.

After generalizing the protocol of Ref.[42] into the case of collective-rotation noise by using the logical Bell states in Eq.(4) to replace the original Bell states used in Ref.[42], it is also easy to find out that the above protocol is highly related to both the second protocol of Ref.[56] and the collective-rotation noise version of the protocol in Ref.[42]. In fact, the above protocol can also be regarded as the combination of these two protocols.

## 4  Security analysis

Without loss of generality, we may consider taking the AQD protocol against the collective-dephasing noise to conduct the security analysis.

(1) Analysis on the information leakage problem

The information leakage problem means that anyone else can obtain partial of classical information just from the public announcement without taking any active attacks. As each two adjacent message logical Bell states $(A_{2n-1}, B_{2n-1})$ and $(A_{2n}, B_{2n})$ are made in the same state by Alice in Step 1, Bob can directly know the initial state of $(A_{2n-1}, B_{2n-1})$ from that of $(A_{2n}, B_{2n})$ after Alice sends both $S'_{EA}$ and $S'_{EB}$ to Bob. Consequently, it is unnecessary for Alice to publish the initial state of $(A_{2n-1}, B_{2n-1})$ to Bob. As a result, Eve has no way to know the initial state of $(A_{2n-1}, B_{2n-1})$. In this case, as to Eve, the measurement result of $(\Omega_{t_d l_d} \Omega_{i_d j_d} A'_d, B'_d)$ involves sixteen kinds of combinations of Alice and Bob's logical unitary operations, which contain $-\sum_{i=1}^{16} p_i \log_2 p_i = -16 \times \frac{1}{16} \log_2 \frac{1}{16} = 4$ bit information from the viewpoint of Shannon's information theory [62]. This amount of information is just equal to the number of encoded classical bits from Alice and Bob. Therefore, no information is leaked out. Apparently, $(A_{2n}, B_{2n})$ acts as the auxiliary logical Bell state helping overcome the information leakage problem, similar to the protocols of Ref.[56].

(2) Analysis on Eve's active attacks

①The impersonation attack

Similar to the protocol of Ref.[42], there are two cases of impersonation attack here.

The first case is that in order to obtain Alice's classical bits, Eve impersonates Bob to communicate with Alice. However, as Eve has no knowledge about $K$ which is used for choosing the measurement bases of sample logical qubits, her random measurements on the sample logical qubits in $S'_{EB}$ cannot produce the genuine $R_{S_{EB}}$. Therefore, Eve can be detected when Alice authenticates the identity of Bob in Step 2 by evaluating the correlation between $R_{S_{EA}}$ and false $R_{S_{EB}}$.

The second case is that in order to obtain Bob's classical bits, Eve impersonates Alice to communicate with Bob. Eve prepares fake $S'_{EB}$, $S'_{EA}$, $S_{OB}$ and $S_{OA}$ beforehand, and sends fake $S'_{EB}$, $S'_{EA}$ and $S_{OB}$ to Bob in Steps 1, 2 and 4, respectively. Then, Bob measures the sample logical qubits in fake $S_{OB}$ with $K$ and obtains a false $R_{S_{OB}}$. Without knowing $K$, Eve randomly measures the corresponding sample logical qubits in fake $S_{OA}$ and obtains a false $R_{S_{OA}}$. Apparently, the correlation between false $R_{S_{OA}}$ and false $R_{S_{OB}}$ is absonant. Therefore, Eve can be detected when Bob authenticates the identity of Alice in Step 4 by evaluating the correlation between false $R_{S_{OA}}$ and false $R_{S_{OB}}$.

②The man-in-the-middle attack

In order to obtain Alice and Bob's classical bits, Eve launches the man-in-the-middle attack by establishing two independent communications with them [42]. Eve prepares fake $S'_{EB}$, $S'_{EA}$, $S_{OB}$ and $S_{OA}$, and replaces all genuine logical qubits traveling between Alice and Bob with them. As analyzed above, Eve will be detected during the mutual authentication between Alice and Bob because she cannot always prepare the genuine initial states and measurement results.

③The modification attack

The aim of modification attack from Eve is to make the participants obtain the wrong classical bits [42]. Similar to the protocol of Ref.[42], two cases of modification attack exist here.

The first case is that after modifying $S''_{OA}$ sent from Alice to Bob in Step 5, Eve sends a modified one to Bob. As $S''_{OA}$ includes $h(m_A)$, the modification attack from Eve will be detected by Bob during the authentication process of Alice's classical bits in Step 6, even if only one logical qubit in $S''_{OA}$ is modified.

The second case is that after modifying the measurement result of $\{(\Omega_{t_1 l_1} \Omega_{i_1 j_1} A'_1, B'_1), (\Omega_{t_2 l_2} \Omega_{i_2 j_2} A'_2, B'_2), \cdots, (\Omega_{t_{3N/4} l_{3N/4}} \Omega_{i_{3N/4} j_{3N/4}} A'_{3N/4}, B'_{3N/4})\}$ (denoted as $R_{msg}$) sent from Bob to Alice in Step 7, Eve sends a modified one to Alice. As $R_{msg}$ includes $h(m_B)$, the modification attack from Eve will be detected by Alice during the authentication process of Bob's classical bits in Step 7.

It can be concluded that the hash operations make the modification attack from Eve detectable.

④The Trojan horse attacks

There are two types of Trojan horse attacks, i.e., the invisible photon eavesdropping [63] and the delay-photon Trojan horse attack [64-65]. In order to launch the invisible photon eavesdropping attack, Eve inserts an invisible photon with an illegitimate wavelength in each quantum signal sent from the sender to the receiver, based on the fact that the single photon detector in the receiver's hand is only sensitive to the photons with a special wavelength [63]. The effective way to defeat the invisible photon eavesdropping attack lies in that the receiver inserts a filter in front of his devices to filter out the photon signal with an illegitimate wavelength before he deals with it [65-66]. The main idea of the delay-photon Trojan horse attack from Eve is to insert a spy photon in a legitimate quantum signal with a delay time, utilizing the fact that the timing for a time window of the optical device has a finite accuracy [64-66]. The effective way to defeat the delay-photon Trojan horse attack lies in that the receiver should use a photon number splitter (PNS:50/50) to split each sample quantum signal into two pieces and measure the signals after the PNS with proper measuring bases [65-66]. If the multiphoton rate is unreasonably high, the existence of this attack will be discovered so that the communication is halted.

## 5 Discussions

(1) The information-theoretical efficiency

Cabello's information-theoretical efficiency [67] is defined as $\eta = b_s/(q_t + b_t)$, where $b_s$, $q_t$ and $b_t$ are the expected secret bits received, the qubits used and the classical bits exchanged between two participants, respectively. Apparently, in the proposed AQD protocol against the collective-dephasing (collective-rotation) noise, after ignoring the quantum resource and the classical resource used for security checks, $(A_d', B_d')$ and its even adjacent logical Bell state can be used to transmit $(i_d, j_d)$ and $(t_d, l_d)$ with two bits consumed on the transmission of the measurement result of $(\Omega_{t_d l_d} \Omega_{i_d j_d} A_d', B_d')$ from Bob to Alice via the classical channel. Accordingly, $b_s = 4$, $q_t = 8$ and $b_t = 2$. Consequently, the information-theoretical efficiency of the proposed AQD protocol against the collective-dephasing (collective-rotation) noise is $\eta = \frac{4}{8+2} \times 100\% = 40\%$.

(2) Comparisons with previous fault tolerant QD protocols

The comparisons between the proposed protocols and those previous fault tolerant QD protocols [53-61] are made here. The comparison results are listed in Table 3, concentrated on the four aspects including the initial quantum resource, the quantum measurement, the information-theoretical efficiency, the information leakage problem and the authentication function. Apparently, only the proposed protocols have the function of identity authentication under the assumption that the information transmitted through the classical channel is eavesdroppable and modifiable. However, similar to the protocols of Ref.[56], the proposed protocols have to prepare two adjacent message logical Bell states in the same state, which enhances the difficulty of experimental implementation.

Table 3  Comparisons with previous fault tolerant QD protocols

|  | Initial quantum resource | Quantum measurement | Information-theoretical efficiency | Information leakage problem | Authentication function |
|---|---|---|---|---|---|
| Ref.[53] | Product states of two original Bell states | Bell state measurements | 40% | No | No |
| Ref.[54] | Logical qubits | Single-photon measurements | 33.3% | No | No |
| Ref.[55] | Logical qubits and single photons | Single-photon measurements | 50% | No | No |
| Ref.[56] | Logical Bell states | Bell state measurements | 40% | No | No |
| Ref.[57] | Nearly Logical qubits | Single-photon measurements | Nearly 66.7% | No | No |
| Ref.[58] | Logical Bell states | Bell state measurements | 40% | No | No |
| Ref.[59] | logical Bell states | Bell state measurements | 40% | No | No |
| Ref.[60] | Logical Bell states | Bell state measurements | 50% | No | No |
| Ref.[61] | Three-qubit entangled states | Single-photon measurements | 50% | No | No |
| The proposed protocols | Logical Bell states | Bell state measurements | 40% | No | Yes |

In addition, it is necessary to further emphasize the differences between the proposed protocols and the protocols of Ref.[60]. The protocols of Ref.[60] each belongs to the kind of quantum secure direct dialogue (QSDD) [37], which requires no additional classical communication for decoding and can accomplish the direct dialogue between two participants. Therefore, the protocols of Ref.[60] each can be considered as the integration of two QSDCs. However, although Bob can decode out Alice's classical bits directly, in order to make Alice able to decode out Bob's classical bits, the proposed protocols each needs the transmission of the measurement result of $(\Omega_{t_d l_d} \Omega_{i_d j_d} A_d', B_d')$ from Bob to Alice via the classical channel. Therefore, the proposed protocols each can be regarded as the combination of QSDC and deterministic secure quantum communication (DSQC) [68-78] rather than the integration of two QSDCs. DSQC is another kind of quantum secret communication which differs from QSDC in that it needs a separate classical communication from the sender to help the receiver decode out the sender's secret messages. It can be concluded that to some extent, the proposed protocols and the protocols of Ref.[60] belong to different types of quantum secret communication. In fact, those previous fault tolerant QD protocols in Refs.[53-59,61] each can also be regarded as the combination of QSDC and DSQC due to the need of an additional classical communication for decoding.

On the other hand, in each of the protocols in Ref.[60], two participants can not only speak to each other either simultaneously or sequentially but also transmit secret messages of different lengths to each other in one round communication. Therefore, the protocols of Ref.[60] are much more flexible than the proposed protocols. However, unlike the proposed protocols, they do not have the function of identity authentication under the assumption that the information transmitted through the classical channel is eavesdroppable and modifiable.

(3) The length of $K$ and its distribution

The key $K$ pre-shared privately between Alice and Bob is used for choosing the measurement bases of sample logical qubits. Alice totally has $3N/2$ classical bits $\{(i_1,j_1)(i_2,j_2)\cdots(i_d,j_d)\cdots(i_{3N/4},j_{3N/4})\}$ while Bob totally has $3N/2$ classical bits $\{(t_1,l_1)(t_2,l_2)\cdots(t_d,l_d)\cdots(t_{3N/4},l_{3N/4})\}$. Therefore, it totally only needs $3N/4$ message logical Bell states for encoding both Alice and Bob's classical bits. In each proposed protocol, Alice prepares $2N$ message logical Bell states $\{(A_1,B_1),(A_2,B_2),\cdots,(A_{2n-1},B_{2n-1}),(A_{2n},B_{2n}),\cdots,(A_{2N-1},B_{2N-1}),(A_{2N},B_{2N})\}$, where only the odd ones are directly used for encoding. It can be concluded that $N-3N/4=N/4$ odd message logical Bell states are used as the sample logical Bell states. Therefore, the length of $K$ is $N/4$. Actually, the length of $K$ dynamically changes along with both the number of message logical Bell states Alice prepares and the amount of two participants' classical bits. Concretely speaking, the length of $K$ is equal to half of the number of message logical Bell states Alice prepares minus a quarter of the amount of two participants' classical bits.

It is well known that quantum key distribution (QKD) [79] aims to establish an unconditionally secure key between two remote users through the transmission of quantum signal. There are some good quantum key distribution (QKD) protocols [79-81] which have been proved to have unconditional security [82-83]. Therefore, Alice can pre-share $K$ with Bob through the QKD protocols of Refs.[79-81].

## 6  Conclusion

In this paper, two fault tolerant AQD protocols against the collective-dephasing noise and the collective-rotation noise are proposed, respectively, by employing logical Bell states as the quantum resource, where the key for choosing the measurement bases of sample logical qubits is pre-shared privately between two participants. Compared with the previous QD protocols [12-24,28-40,42,53-61], the great merit of the two proposed protocols is that each of them can accomplish the mutual identity authentication and the dialogue between two participants simultaneously and securely over one kind of collective noise channels under the assumption that the information transmitted through the classical channel is eavesdroppable and modifiable. Their other characteristics are:

(1) With respect to the quantum measurement, the Bell state measurements rather than the four-qubit joint measurements are adopted;

(2) On the aspect of information leakage problem, the method of direct transmission of auxiliary logical Bell states is employed to avoid it;

(3) As to Eve's active attacks, the impersonation attack, the man-in-the-middle attack, the modification attack and the Trojan horse attacks all are detectable.

**Acknowledgements**

The author would like to thank the anonymous reviewers for their valuable suggestions that help enhancing the quality of this paper. Funding by the National Natural Science Foundation of China (Grant Nos.61402407, 11375152) is also gratefully acknowledged.

**Compliance with ethical standards**

Conflict of interest: The author declares that he has no conflict of interest.